\documentclass[aps,pra,reprint,groupedaddress]{revtex4-1}

\usepackage{amssymb}
\usepackage{bm}

\begin{document}


\title{Zero range potential for particles interacting via Coulomb potential: application to electron positron annihilation}

\author{S.L. Yakovlev}
\email[E-mail: ]{yakovlev@cph10.phys.spbu.ru}
\affiliation{Department of Computational Physics, St Petersburg State University, 198504, St Petersburg, Russia }
\author{V.A. Gradusov}
\email[E-mail: ]{vitaly.gradusov@gmail.com}
\affiliation{Department of Computational Physics, St Petersburg State University. 198504, St Petersburg, Russia }

\author{}
\affiliation{}



\begin{abstract}
The zero range potential is constructed for a system of two particles interacting via the Coulomb potential.
The singular part of the asymptote  of the wave function at the origin which is caused by the common effect of the zero range potential singularity and of the Coulomb potential  is explicitly
calculated  by using the Lippmann-Schwinger type integral equation. The singular pseudo potential is constructed from the requirement that it enforces the solution to the Coulomb Schr\"odinger equation to possess   the calculated asymptotic behavior at the origin.
This pseudo potential is then used for constructing a model of the imaginary absorbing potential for the positron electron system. This potential allows to treat the annihilation process in positron electron collisions on the basis of the non relativistic Schr\"odinger equation.
The functional form of the
pseudo potential constructed in this paper is analogous to the well known Fermi-Breit-Huang pseudo potential.
The generalization of the optical theorem on the case of the imaginary absorbing potential in presence of the Coulomb force is given in terms of the partial wave series.
\end{abstract}


\maketitle


\section{Introduction}

The Dirac-delta potential was introduced by Fermi to describe the interaction of neutrons with atomic nucleus~\cite{fermi}. Already at that time it was realized  that only perturbative  treatment of such a potential is possible~\cite{breit} due to the strong singularity of the delta function in three dimensional.

 One of the possible way to go beyond the perturbative treatment is the use of the concept of zero range potentials~\cite{breit, minlos-f, demkov}. In this approach we impose the singular boundary condition
\begin{equation}
\phi(\bm{r},\bm{k})\sim \frac{\alpha_s}{r}+\beta_s
\label{bc0}
\end{equation}
as $r\to 0$.
Here $\phi(\bm{r},\bm{k})$ is the solution of the Schr\"odinger equation
\begin{equation}
\left[-\Delta_{\bm{r}}+V(\bm{r})-k^2\right]\phi(\bm{r},\bm{k})=0,
\label{sch-e}
\end{equation}
where $\Delta$ stands for the Laplacian, bold letters are used for vectors in $R^3$ like $\bm{r}$ and $\bm{k}$, and $r=|\bm{r}|$ and
$k=|\bm{k}|$.
The potential $V(\bm{r})$ here should be of a short range type, i.e. it should be less singular than $1/r^2$
at the origin  and should vanish  faster than the Coulomb potential $1/r$ as $r\to \infty$ ~\cite{albev}.
The parameters $\alpha_s$ and $\beta_s$ are not independent and should be correlated in such a way that the wave function
$\phi(\bm{r,\bm{k}})$ is the solution of Eq.\ (\ref{sch-e}) when $r\to 0$. As a result,
 the only independent parameter $a_s$ of the model
 can be chosen as $a_s=-\alpha_s/\beta_s$. In the case of $V(\bm{r})=0$ this parameter is the scattering length.

To enforce the boundary condition in an alternative approach ~\cite{blatt,hua57} an additional term is introduced into the Schr\"dingier equation
\begin{equation}
\left[-\Delta_{\bm{r}}+V(\bm{r}) -k^2\right]\phi(\bm{r},\bm{k})+ \lambda \delta(\bm{r})\beta_s = 0,
\label{sch-e-delta}
\end{equation}
where  $\lambda = 4\pi a_s$. Now  $\beta_s$ can be obtained as the limit
\begin{equation}
\beta_s= \lim_{r\to 0} \frac{\mbox d}{{\mbox d} r}r\phi(\bm{r}, \bm{k}).
\label{beta}
\end{equation}
This allows us to rewrite  (\ref{sch-e-delta})   as~ \cite{hua57}
\begin{equation}
\left[-\Delta_{\bm{r}}+V(\bm{r}) + \lambda W_s(\bm{r})-k^2\right]\phi(\bm{r},\bm{k})=0,
\label{sch-e-W}
\end{equation}
where
\begin{equation}
W_s(\bm{r})= \delta( {\bm{r}})\frac{\mbox d}{{\mbox d} r}r~
 \label{W}
\end{equation}
is a kind of "pseudo" potential.
We note that all above expressions can be derived \cite{albev} from the Lippmann-Schwinger equation
\begin{equation}
\phi(\bm{r},\bm{k})=\phi_0(\bm{r},\bm{k}) - \lambda\int \mbox{d} \bm{r}' \, G(\bm{r},\bm{r}',k^2+i0)W_s(\bm{r}')\psi(\bm{r}',\bm{k}),
\label{LS}
\end{equation}
where $\phi_0(\bm{r},\bm{k})$ is the regular solution to the equation (\ref{sch-e-delta}) with $\lambda=0$ and $G(\bm{r},\bm{r}',k^2+i0)$ is the Green's function for the Hamiltonian $-\Delta_{\bm{r}}+V(\bm{r})$.
The explicit form of $W$ given in Eq.\ (\ref{W}) is valid only for short-range potentials and it should be modified for long-range potentials like the Coulomb potential \cite{kuper88}.

Zero range potentials are the mathematically correct tools for describing contact interactions. However, in applications  the delta function is often used as a potential of contact interaction.
A recent example of such an application of the delta function potential is its use in describing the annihilation of
positrons in atomic systems~\cite{iva00}. The loss of particles in positron-hydrogen scattering due to the $e^+e^-$ annihilation  can be simulated by an imaginary absorption potential in the positron-electron subsystem.
Arguments from perturbative QED  \cite{humber} suggest a three-dimensional delta-function potential  \cite{iva00}.  In a series of calculations of positron hydrogen scattering  \cite{iva00,iga02,iga03,yam03} this delta functional potential was smoothed out and the  delta function was replaced by Gaussian with finite width. Thus the concept of zero-range potentials was not used. Nevertheless, a mathematically sound formulation of how the three dimensional zero range potentials in the Coulombic systems can be used is of definite demand. A way for such a formulation was outlined in Ref.~\cite{yak07}.

In this  paper we continue developing the zero range potential concept in the presence of Coulomb interaction.
We give here the detailed derivation of the pseudo potential for the two body Coulomb Hamiltonian. We show that, as in the short range case, the Schr\"odinger equation for this potential can be solved analytically with the help  of a
Lippmann-Schwinger--type integral equation. We derive from this solution the exact asymptote of the wave function at the origin and then the corresponding pseudo potential is constructed. We show that, if this potential is treated perturbatively, it can be taken in the simple form of the delta function as it was used in \cite{iva00} for the  positron electron case.
Finally, we apply the constructed zero range potential to the description of the positron electron annihilation by introducing the imaginary absorbing pseudo potential. With this potential we solve the bound state problem deriving the positronium spectrum as well as the scattering problem calculating the positron electron wave function and the scattering amplitude. The annihilation cross section corresponds to the loss of the flux 
due to the imaginary absorbing potential. It is calculated by using a generalized optical theorem which we have modified in such a way it can be applicable to
the case of Coulomb potential.

\section{\label{potcon}Pseudo potential for Coulomb Hamiltonian}

In this section we construct the zero range potential and the corresponding  pseudo potential for the  Coulomb Hamiltonian $H_{\cal C}$
\begin{equation}
H_{\cal C}= -\Delta_{\bm{r}}+\mbox{n}/r,
\label{HC}
\end{equation}
where
\begin{equation}
\mbox{n}=\epsilon \frac{2\mu e^2 Z_1 Z_2}{\hbar^2}
\label{en}
\end{equation}
 with $\epsilon=\pm 1$ for the repulsive and attractive case, respectively.
The Coulomb wave function $\psi_{\cal C}(\bm{r},\bm{k})$ satisfies the equation
\begin{equation}
(H_{\cal C}-k^2)\psi_{\cal C}(\bm{r},\bm{k})=0.
\label{Coul SE}
\end{equation}
The explicit form of $\psi_{\cal C}(\bm{r},\bm{k})$ reads \cite{messiah}
\begin{equation}
\psi_{\cal C}(\bm{r},\bm{k})= \Gamma(1+i\eta)e^{-\pi\eta/2}e^{i \bm{r}\cdot\bm{k}} M(-i\eta,1;i[rk-\bm{r}\cdot\bm{k}]),
\label{Cwf}
\end{equation}
where $\eta=\mbox{n}/(2k)$,  $\Gamma(z)$ is the gamma function and $M(a,b;z)$ is the regular Kummer (or confluent hypergeometric) function \cite{abram}.

Now we construct the zero-range potential and the corresponding pseudo potential along the scheme given
by Eqs.\ (\ref{bc0}-\ref{LS}). The modified Schr\"odinger equation reads
\begin{equation}
\label{initeq}
\left[ H_{\cal C}-k^2\right]\psi(\bm{r},\bm{k})=-\lambda W(\bm{r})\psi(\bm{r},\bm{k})
\end{equation}
and $W$ is such that
\begin{equation}
W(\bm{r})\psi(\bm{r},\bm{k})=\delta(\bm{r})\beta.
\label{WC}
\end{equation}
The parameter $\beta$  can be determined from the solution of the Eq.\ (\ref{initeq}). This solution, as in the short range case, can be obtained by solving the Lippmann-Schwinger equation
\begin{equation}
\psi(\bm{r},\bm{k})= \psi_{\cal C}(\bm{r},\bm{k})-\lambda \int\mbox{d}\bm{r}'\, G_{\cal C}(\bm{r},\bm{r}',k^2+i0)W(\bm{r}')\psi(\bm{r}',\bm{k}).
\label{CLS}
\end{equation}
Here $G_{\cal C}$ stands for the Coulomb Green's function. In view of Eq.\ (\ref{WC}) the integration on the right hand side of
Eq.\ (\ref{CLS}) immediately leads to the representation
\begin{equation}
\label{Cintrep}
\psi(\bm{r},\bm{k})=\psi_{\cal C}(\bm{r},\bm{k})-\lambda G_{\cal C}(\bm{r},0,k^2+i0)\beta.
\end{equation}
From the explicit form of the Coulomb Green's function given in~\cite{hostl64} one can derive the following representation in terms of the  Whittaker function 
~\cite{abram}
\begin{equation}
\label{coulgf}
G_{\cal C}(\bm{r},0,k^{2}+i0)=\frac{1}{4\pi r}\Gamma(1+i\eta)W_{-i\eta;\frac{1}{2}}(-2ikr)~.
\end{equation}
This representation  is the basis for computing
the asymptotic expansion of $\psi(\bm{r},\bm{k})$ as $r\to 0$. The Coulomb wave function is regular at $r=0$ and can be expanded as
\begin{equation}
\psi_{\cal C}(\bm{r},\bm{k})=\Gamma(1+i\eta)e^{-\pi\eta/2}[1 + {\cal O}(r)].
\label{psiC0}
\end{equation}
The Coulomb Green's function $G_C(\bm{r},0,k^{2}+i0)$ is singular as $r\to0$.
We will determine the explicit form of this singularity by using the representation (\ref{coulgf}).
The Whittaker function can be represented by the irregular Kummer function~\cite{abram}
\begin{equation}
\label{wu}
W_{\gamma;\mu}(z)=e^{-\frac{1}{2}z}z^{\frac{1}{2}+\mu}U(
1/2+\mu-\gamma,1+2\mu,z),
\end{equation}
which is valid for $-\pi<\arg{z}<\pi$.  With $\gamma=-i\eta$ it results
\begin{equation}
W_{-i\eta;\frac{1}{2}}(-2ikr)=-2ikr e^{ikr}U(1+i\eta,2,-2ikr).
\label{W-U}
\end{equation}
We use following representation of the irregular Kummer function expansion \cite{abram} for integer $m$
\begin{widetext}
\begin{eqnarray}
 U(a,m+1,z)= \frac{(-1)^{m+1}}{m!\Gamma(a-m)}\left[M(a,m+1;z)\ln z \right. \nonumber \\
+ \left. \sum_{l=0}^{\infty}\frac{(a)_{l}}{l!(m+1)_{l}}z^{l}\left\{\psi(a+l)-\psi(1+l)-\psi(1+m+l)\right\} \right] 
 + \frac{(m-1)!}{\Gamma(a)}z^{-m}M(a-m,1-m;z)_{m}. \qquad
\label{U ser}
\end{eqnarray}
\end{widetext}
Here $-\pi<\arg{z}<\pi$, $(a)_{l}=\Gamma(a+l)/\Gamma(a)$ is the Pochhammer symbol, $\psi(z)=\Gamma'(z)/\Gamma(z)$ is the digamma function and $M(a,b;z)_m$
is the truncated regular Kummer function defined by the polynomial
\begin{equation}
\label{1f1def}
M(a,b;z)_m= 
\sum_{l=0}^{m}\frac{(a)_{l}}{l!(b)_{l}}z^{l}.
\end{equation}
The regular Kummer function $M(a,b;z)$ is given in terms of $M(a,b;z)_m$ by the limit
$$
M(a,b;z)=\lim_{m\to \infty}M(a,b;z)_{m}.
$$
Using now (\ref{U ser}) for the $U$ factor in (\ref{W-U}) with appropriate arguments and keeping the leading terms
we arrive at the following asymptotic representation for that $U$ factor  as $r\to 0$
\begin{widetext}
\begin{eqnarray}
\label{uasym}
U(1+i\eta,2,-2ikr)=
-\frac{1}{2ik}\frac{1}{\Gamma(1+i\eta)}\left[
1/r+2k\eta\log r\right] 
+ \frac{i\eta}{\mbox{n}\Gamma(1+i\eta)}
\left[4\pi C(k)-ik \right] 
 + {\cal{O}}(r\log r).
\end{eqnarray}
\end{widetext}
Here $C(k)$ is given by
\begin{equation}
C(k)=\frac{ik}{4\pi}+\frac{\mbox{n}}{4\pi}\left[ \log(-2ik)+\psi\left(1+i\eta\right)+2\gamma_{0}-1 \right] ,
\label{C(k)}
\end{equation}
where $\gamma_0$ is the Euler-Mascheroni constant \cite{abram}.
Substituting~(\ref{W-U})~and~(\ref{uasym}) into~(\ref{coulgf}) we obtain the required Coulomb Green's function as $r\to 0$
\begin{equation}
\label{gfasym}
G_{\cal C}(\bm{r},0,k^{2}+i0) = \frac{1}{4\pi}\left[
1/r+\mbox{n}\log r\right]+C(k)+{\cal{O}}(r\ln r).
\end{equation}
With these results  introduced into~(\ref{Cintrep}),  
we end up with the following asymptotic expansion for the wave function as $r\to 0$
\begin{equation}
\label{wfas}
\psi(\bm{r},\bm{k}) = \frac{-\lambda \beta}{4\pi}\left[
1/r+\mbox{n}\log r \right]+\psi_{\text{reg}}+{\cal{O}}(r\log r),
\end{equation}
where
$\psi_{\text{reg}}=\Gamma(1+i\eta)e^{-\pi\eta/2}-\lambda \beta C(k)$ is the regular part, which can be idetified as $\beta=\psi_{\text{reg}}$.
The asymptotic expansion of the wave function as $r\to 0$  now takes the final form
\begin{equation}
\psi(\bm{r},\bm{k}) =\frac{\alpha}{4\pi}\left[
1/r+\mbox{n}\log r \right]+\beta+{\cal{O}}(r\log r),
\label{psi BC}
\end{equation}
and the constants  are given by
\begin{eqnarray}
\alpha&=&-\lambda \beta,  \label{alph} \\  
\beta&=&\frac{\Gamma(1+i\eta)e^{-\pi\eta/2}}{1+\lambda C(k)}.
\label{bet}
\end{eqnarray}
As in the short range case, the only independent  parameter $\lambda$ now determines the zero range potential. Similar to the short range case $\lambda$ is related to the ratio of parameters $\alpha$ and $\beta$ by $\lambda=-\alpha/\beta$.

At the last stage of our construction let us represent $\beta$ in terms of the wave function. 
Introducing a new variable
$$
u=\frac{r}{1+\mbox{n} r\log r}
$$
it is easy to obtain by direct calculation the following result
\begin{equation}
\lim_{r\to 0}\frac{\mbox{d}}{\mbox{d}u}u\,\psi(\bm{r},\bm{k})=\beta.
\label{lim u}
\end{equation}
This formula allows us to rewrite the pseudo potential $W$ in the form
$$
W(\bm{r})=\delta(\bm{r})
\frac{\mbox{d}}{\mbox{d}u}u~, 
$$
and even more explicitly in the final form
\begin{equation}
\label{ppot}
W(\bm{r})=\delta(\bm{r})\frac{(1+\mbox{n} r\log r)^{2}}{1-\mbox{n} r}\frac{\mbox{d}}{\mbox{d}r}\frac{r}{1+\mbox{n}r\log r}\, .
\end{equation}
We can see that in the non-Coulomb case, when $n=0$, we recover the short-range result of Eq.\ (\ref{W}).

\section{Structure of the Green's Function}

In this section we construct the Green's function for the Hamiltonian $H_{\cal C}+\lambda W(\bm{r})$.
It is  the solution of the equation
\begin{equation}
\label{gfdef}
\left[ H_{\cal C}+ \lambda W  -z \right]G(\bm{r},\bm{x},z)=\delta(\bm{r}-\bm{x}).
\end{equation}
The Green's function is symmetric with respect to $\bm{r}$ and $\bm{x}$ and obeys the boundary condition as $r\to 0$
\begin{equation}
\label{gfbound}
G(\bm{r},\bm{x},z)=\frac{A(\bm{x},z)}{4\pi}\left[ 
{1}/{r}+\mbox{n}\log r \right]+B(\bm{x},z)+{\cal O}(r\log r)~.
\end{equation}
Like in Eq.\ (\ref{WC}), the action of the quasi potential on the Green's function  is given by
$$
W(\bm{r})G(\bm{r},\bm{x},z)=\delta(\bm{r})B(\bm{x},z).
$$
Thus, we can combine (\ref{gfdef}) and (\ref{gfbound}) to obtain
\begin{equation}
\left[ H_{\cal C} -z \right]G(\bm{r},\bm{x},z)=\delta(\bm{r}-\bm{x}) - \lambda \delta(\bm{r})B(\bm{x},z).
\end{equation}
By inverting  the operator $H_{\cal C}-z$  with the help of the Coulomb Green's function, and performing the integration with the  the delta function, we arrive at the representation
\begin{equation}
\label{gfe}
G(\bm{r},\bm{x},z)= G_{\cal C}(\bm{r},\bm{x},z) - \lambda G_{\cal C}(\bm{r},0,z) B(\bm{x},z).
\end{equation}
Taking the limit when $r\to 0$ for a fixed nonzero $\bm{x}$  with the use of~(\ref{gfasym}) we arrive at the
expression
\begin{eqnarray}
G(\bm{r},\bm{x},z)  =  \frac{-\lambda B(\bm{x},z)}{4\pi}\left[ 
{1}/{r}+\mbox{n}\log r \right] \nonumber \\
  + G_{\cal C}(0,\bm{x},z)-\lambda B(\bm{x},z)C(\sqrt{z})+{\cal O}(r\log r).
\end{eqnarray}
Comparing  this asymptotic form with the definition~(\ref{gfbound})
we find the expressions for
 $A(\bm{x},z)$ and $B(\bm{x},z)$
\begin{eqnarray}
A(\bm{x},z) & = &  G_{\cal C}(0,\bm{x},z)\frac{-\lambda}{1+\lambda C(\sqrt{z})} \nonumber \\
B(\bm{x},z) & = & G_{\cal C}(0,\bm{x},z)\frac{1}{1+\lambda C(\sqrt{z})}.
\end{eqnarray}
By substituting $B(\bm{x},z)$ into~(\ref{gfe}), we obtain the final representation for the Green's function in the form
\begin{eqnarray}
G(\bm{r},\bm{x},z) & =&  G_{\cal C}(\bm{r},\bm{x},z)  \nonumber \\
& &-\lambda G_{\cal C}(\bm{r},0,z) \frac{1}{1+\lambda C(\sqrt{z})} G_{\cal C}(0,\bm{x},z).\ \ \
\label{GF}
\end{eqnarray}
As a straightforward consequence of this representation, the  equation
\begin{equation}
\label{disper}
1+\lambda C(\sqrt{z})=0
\end{equation}
determines the poles of the Green's function, or in other words, the spectrum of eigenvalues of the Hamiltonian $H_{\cal C}+\lambda W(\bm{r})$.

The Eq. (\ref{GF}) can be written in the operator form
\begin{equation}
G(z)=G_{\cal C}(z)-G_{\cal C}(z){\cal T}(z)G_{\cal C}(z),
\label{GT}
\end{equation}
were the ${\cal T}$ matrix  is defined by the kernel
\begin{equation}
{\cal T}(\bm{r},\bm{x},z) = \lambda \frac{\delta(\bm{r})\delta(\bm{x})}{1+\lambda C(\sqrt{z})}.
\label{T}
\end{equation}
If now we formally introduce the Coulomb T-matrix by the representation of the Coulomb Green's function
\begin{equation}
G_{\cal C}(z)=G_0(z) - G_0(z)T_{\cal C}(z)G_0(z),
\label{GCT}
\end{equation}
where $G_0(z)=(-\Delta - z)^{-1}$, then the Eq. (\ref{GT}) can further be rewritten in the standard form
\begin{equation}
G(z)=G_0(z)-G_0(z)T(z)G_0(z).
\label{GG0}
\end{equation}
Here the T-matrix $T(z)$ for Coulomb plus zero range potentials  is given by the explicit expression
\begin{equation}
T(z)= T_{\cal C}(z) + [I-T_{\cal C}(z)G_0(z)]{\cal T}(z)[I-G_0(z)T_{\cal C}(z)].
\label{TTC}
\end{equation}
This formula reproduces the conventional representation for the T-matrix in the case of the Coulomb plus short-range interaction which follows from the
standard two potential formalism \cite{messiah}.

\section{Electron positron annihilation potential}

The  formula for the 2$\gamma$ singlet $e^{+}e^{-}$ annihilation  cross section derived from perturbative QED reads~\cite{fra68,humber}
\begin{equation}
\label{qed}
\sigma_{\text{ann}}=\sigma_0 %
Z_{\text{eff}},
\end{equation}
where
$
\sigma_0= \pi r^2_0 (c/v)
$,
$r_0$ is the classical electron radius, $c$ is the speed of light and $v$ is the incident velocity of the positron.
The effective number of electrons $Z_{\text{eff}}$ participating in the annihilation is given by  the integral
\begin{equation}
Z_{\text{eff}} = \int\,d\bm{r}_{1}\,d\bm{r}_{2}\,|\Psi^{0+}(\bm{r}_{1},\bm{r}_{2},\bm{r}_{p})|^{2}\delta(\bm{r}_{1}-\bm{r}_{2}).\ \ \ \
\end{equation}
Here $\Psi^{0+}(\bm{r}_{1},\bm{r}_{2},\bm{r}_{p})$ (with $\bm{r}_1$, $\bm{r}_2$ and $\bm{r}_p$ being the position vectors of the positron, the electron and the proton) is the solution of the positron hydrogen scattering problem, when annihilation is not taken into account, and normalized in such a way that the incident wave has the form
\begin{equation}
\exp(i\bm{k}_1\cdot \bm{r}_{1})\Phi(\bm{r}_{2},\bm{r}_{p}),
\end{equation}
 where $\Phi(\bm{r}_{2},\bm{r}_{p})$ is the hydrogen wave function.

In the Ref.\ \cite{iva00}, to describe the absorption in the positron hydrogen scattering process an imaginary optical potential of the form $igW_{12}$, with $g<0$, was introduced. The optical theorem,   in the first order Born approximation, gives the following expression for the absorption cross section 
\begin{eqnarray}
\label{born}
\sigma_{\text{ann}}^{B} &=& \frac{ (-g)}{k_1}\nonumber \\
&& \times \int\,d\bm{r}_{1}\,d\bm{r}_{2}\,W_{12}(\bm{r}_{1}-\bm{r}_{2}) |\Psi^{0+}(\bm{r}_{1},\bm{r}_{2},\bm{r}_{p})|^{2}.\ \ \
\end{eqnarray}
By comparing the QED expression~(\ref{qed}) with Eq.~(\ref{born}) and taking into account that $\sigma_{\text{ann}}=\sigma_{\text{ann}}^{B}/4$  (the factor $1/4$ comes from the fact that only singlet state is to be taken into account) it was proposed in \cite{iva00} to determine the optical potential
in such a way it has the coupling
constant of the form
\begin{equation}
g=- k_1\sigma_0
\label{Yakovlev Hu g}
\end{equation}
and the coordinate part of the form
\begin{equation}
W_{12}(\bm{r})=\delta(\bm{r}),
\label{Mitroy delta}
\end{equation}
where $\bm{r}=\bm{r}_{1}-\bm{r}_{2}$.

As we have discussed above, such a choice of the coordinate part of the potential leads to troubles in applications since the delta function as a potential can only be treated on the basis of the perturbation theory. Moreover, as it was studied in details in Ref.\
\cite{yak07}, the formula (\ref{born}) is not valid for energies above the positronium formation threshold. In that paper 
it was proposed to use the zero range potential as a remedy for all of these problems.
Thus, the actual positron electron annihilation potential
in the framework of above considerations should be defined as
\begin{equation}
ig W_{12}(\bm{r})= ig W(\bm{r}),
\label{W ann}
\end{equation}
where $W(\bm{r})$ is the pseudo potential of Eq.\ (\ref{ppot}). Consequently, the positron electron Hamiltonian, which simultaneously describes the inter particle dynamics and annihilation, takes the form
\begin{equation}
H = H_{\cal C} + ig W(\bm{r}).
\label{pe H}
\end{equation}

\section{The positronium spectrum}

In this section we study the positronium  spectrum  as the discrete spectrum of the positron-electron Hamiltonian $H_{\cal C} + ig W(\bm{r})$ with annihilation potential ~(\ref{W ann}). Here is the case when
$\epsilon=-1$ in Eq.\ (\ref{en}) and  $\mbox{n}<0$. The spectrum of $H$ is formed by zeros of the denominator in (\ref{GF}) when  $\lambda=ig$. For analyzing solutions of the equation (\ref{disper}) we rewrite it in the form
\begin{equation}
\label{eneq1}
C(\sqrt{z}) = i{g}^{-1},
\end{equation}
where
\begin{eqnarray}
C(\sqrt{z})= \frac{i\sqrt{z}}{4\pi}& \nonumber
\\
+\frac{\mbox{n}}{4\pi}\left[  \log(-2i\sqrt{z})+\psi\left(1+i\frac{\mbox{n}}{2\sqrt{z}}\right)+2\gamma_{0}-1 \right]&.
\label{C(z)}
\end{eqnarray}
Since the actual value of the coupling constant $g$ is small, around $10^{-6}$ eV,  we can seek the solution of Eq.\ (\ref{eneq1}) in the $g\to 0$ limit.
In this limit  the right hand side of (\ref{eneq1}) tends to infinity and so should do the left hand side.
It may happen only due to the digamma function.
Using the well known series expansion of the digamma function~\cite{abram}
\begin{equation}
\label{psi}
\psi(x)=-\gamma_{0}-\frac{1}{x}+\sum_{m=1}^{\infty}\left( \frac{1}{m} - \frac{1/m}{1+x/m} \right)
\end{equation}
we see that if the digamma function argument is equal to $-m$, where $m$ is a non-negative integer, the left hand side of
Eq.\ (\ref{eneq1}) will obey the required property.  Thus, we get the equation that determines the zero order approximation to energy levels
\begin{equation}
1+i\frac{\mbox{n}}{2\sqrt{z}}=-m,\qquad m=0,1,\ldots~.
\end{equation}
Solving it, we recognize the pure Coulomb spectrum
\begin{equation}
z_{N}=-\frac{\mbox{n}^2}{4N^2},\qquad N=1,2,\ldots \,
\end{equation}
in accordance with the obvious expectation when no absorption is present in the Hamiltonian if $g=0$.

As the next stage we will calculate the first order correction  to $z_N$. 
First we represent the exact energy level in the form
\begin{equation}
z'_N=-(\varkappa_{N}+\delta_{N})^{2}, 
\label{z''}
\end{equation}
 where $-\varkappa_{N}^{2}=z_{N}$ is the Coulomb energy and $\delta_{N}$ is the correction of interest due to the presence of the annihilation potential. Obviously, we expect to have $\delta_{N}=O(g)$.
For calculating $\delta_N$ let us use the inverse to the equation~(\ref{eneq1})
\begin{equation}
\label{eneq2}
\frac{1}{C(\sqrt{z})}=
-ig.
\end{equation}
According to the discussion above, the digamma function argument for values of $z$ in the vicinity of $z_N$
tends to a nonpositive integer when $g\to0$, i.e.
\begin{equation}
1+i\frac{\mbox{n}}{2\sqrt{z'_N}}  =  
1-N-\frac{\mbox{n}}{2\varkappa_{N}^{2}}\delta_{N}+{\cal O}(g^{2})
\qquad N =  1,2,\ldots.
\end{equation}
Due to Eq.\  (\ref{psi}) for the leading term of the digamma function  when  $x\to -p$ with integer $p$ we have
\begin{equation}
\psi(x) \sim -\frac{1}{x+p},\qquad p=0,1,\ldots.
\label{diga}
\end{equation}
So, we can conclude that
\begin{equation}
\frac{1}{\psi\left(1+i\frac{\mbox{n}}{2\sqrt{z'_N}}\right)}=\frac{\mbox{n}}{2\varkappa_{N}^{2}}\delta_{N} + {\cal O}(g^{2}), 
\label{diga1}
\end{equation}
where $N=1,2,\ldots$.
Now the left hand side of the equation  (\ref{eneq2}) takes the form
\begin{equation}
\frac{1}{C(\sqrt{z'_N})}=\frac{4\pi}{\mbox{n}\,\psi\left(1+i\frac{\mbox{n}}{2\sqrt{z'_N}}\right)}+{\cal O}(g^{2}).
\label{eqn}
\end{equation}
Considering (\ref{diga1}) we can rewrite this in the  form
\begin{equation}
\frac{2\pi}{\varkappa_{N}^{2}} \delta_{N}+{\cal O}(g^{2})=-ig~,
\label{eqn2}
\end{equation}
and for $\delta_N$ we obtain
\begin{equation}
\delta_{N}=-\frac{ ig\varkappa_{N}^{2}}{2\pi}+{\cal O}(g^{2})=-\frac{ig\mbox{n}^2}{8\pi N^{2}}+{\cal O}(g^{2}). 
\label{delta}
\end{equation}
 Consequently,  the energy levels are given by
\begin{equation}
z'_N=-\varkappa_{N}^2-2\varkappa_{N}\delta_{N}-\delta_{N}^{2}=-\varkappa_{N}^{2}+ig\frac{(-\mbox{n})^3}{8\pi N^3}+{\cal O}(g^2)~
\label{z'}
\end{equation}
with $N=1,2,\ldots$.
So we got the first order correction to energy levels  due to the annihilation potential
\begin{equation}
\label{de}
ig\frac{(-\mbox{n})^3}{8\pi N^3},\qquad N=1,2,\ldots~.
\end{equation}

It is interesting to compare this value with the result of the standard perturbation theory. For the potential $igW(\bm{r})$
the first order energy level  correction is given by the matrix element
\begin{equation}
\label{corr}
\langle \psi_{N\ell m}|igW|\psi_{N\ell m}\rangle = \int\,d\bm{r}\,\psi_{N\ell m}(\bm{r})igW(\bm{r})\psi_{N\ell m}(\bm{r})
,
\end{equation}
where $\psi_{N\ell m}(\bm{r})$ is the Coulomb bound state wave function \cite{messiah}
\begin{eqnarray}
\psi_{N\ell m}(\bm{r})=\sqrt{\frac{(-\mbox{n})^3}{8}}\frac{2}{N^2}\sqrt{\frac{(N-\ell-1)!}{[(N+\ell)!]^3}} \nonumber \\
\times \exp\left\{\frac{\mbox{n}r}{2N}\right\}
\left(\frac{-\mbox{n}r}{N}\right)^{\ell}L_{N-\ell-1}^{2\ell+1}\left(\frac{-\mbox{n}r}{N}\right)Y_{\ell}^{m}(\theta,\phi).
\label{Coulbs}
\end{eqnarray}
The integral on the right hand side of (\ref{corr}) can easily be evaluated to give
\begin{equation}
\label{corr1}
\langle \psi_{Nlm}|igW|\psi_{Nlm}\rangle 
=ig\psi_{Nlm}^2(0).
\end{equation}
Since $\psi_{Nlm}^2(0)$ is non-singular only for s-wave, i.e. for $l=0$,  we have
\begin{eqnarray}
\psi_{N00}^2(0)=\frac{(-\mbox{n})^3}{8\pi N^3}.
\label{psi0}
\end{eqnarray}
This leads us to the final result for the perturbative correction
\begin{equation}
\langle \psi_{N00}|igW|\psi_{N00}\rangle= ig\frac{(-\mbox{n})^3}{8\pi N^3},
\label{perturb}
\end{equation}
which is in complete agreement with our previous result.

\section{Electron-positron scattering}

In this section we consider the positron electron scattering with annihilation. The latter is described by the annihilation
potential $igW(\bm{r})$. The corresponding wave function  $\psi(\bm{r},\bm{k})$ is the solution of the Schr\"odinger equation (\ref{initeq})
with $\lambda = ig$ and it possess the asymptotic behavior as $r\to \infty$
\begin{eqnarray}
\label{scatc}
\psi(\bm{r},\bm{k})  \sim  \exp[ i\bm{k}\cdot \bm{r}+ i\eta \log( kr-\bm{k}\cdot \bm{r} ) ] \nonumber \\
  +  f(\theta) \frac{ \exp[ikr-i\eta\log(2kr)] }{r}\ ,
\end{eqnarray}
where $\theta$ is the scattering angle defined such  that $\cos \theta= \bm{r}\cdot \bm{k}/(rk)$ and  $f(\theta)$ is the scattering amplitude.
As we have shown in Sec.\ \ref{potcon}, this wave function satisfies the singular boundary condition at the origin
\begin{equation}
\label{0bc}
\psi(\bm{r},\bm{k}) = \frac{\alpha}{4\pi} \left[ 1/r + \mbox{n}\log r \right] +\beta +{\cal O}(r\log r)
\end{equation}
with $\alpha/\beta=-i{g}$. 
It is also the solution of the integral equation
(\ref{CLS}) with $\lambda=ig$, and consequently due to (\ref{Cintrep}), (\ref{coulgf}) and (\ref{bet})
is can be written as
\begin{equation}
\label{scatwf}
\psi(\bm{r},\bm{k})=\psi_{\cal C}(\bm{r},\bm{k})  - ig\frac{  \Gamma^2(1+i\eta)
e^{-\pi\eta/2}}{ 1+igC(k)} \frac{ W_{-i\eta;\frac12}(-2ikr) }{4 \pi r}\, .
\end{equation}
By examining the asymptotic behavior  of this wave function as $r\to \infty$ we can determine the scattering amplitude $f(\theta)$.
The asymptotic expansion can be obtained from (\ref{scatwf}) by using the well known asymptotic representations
for the Coulomb wave function $\psi_{\cal C}(\bm{r},\bm{k})$ and the Whittaker function.
The first term in (\ref{scatwf}) has the form of (\ref{scatc}) with the Coulomb scattering amplitude \cite{messiah}
\begin{equation}
f_{\cal C}(\theta)=-\frac{\eta}{2k\sin^2\theta/2}\exp\{-i\eta\log( \sin^2\theta/2) + 2i\sigma_0\}.
\label{f_C}
\end{equation}
The Coulomb phase shift here is given by $\sigma_0=\arg \Gamma(1+i\eta)$.
The asymptotics of the Whittaker function as $r\to \infty$ reads  \cite{abram}
\begin{equation}
W_{-i\eta,\frac12}(-2ikr) = 
e^{-\pi\eta/2} e^{ikr - i\eta\log(2kr)} + {\cal O}(1/r).
\end{equation}
Introducing these asymptotic representations into (\ref{scatwf}) we recover the asymptotic expansion (\ref{scatc}) and as the result we obtain the exact representation for the scattering amplitude $f(\theta)$
\begin{equation}
f(\theta)= f_{\cal C}(\theta) + f'(\theta),
\label{ampl}
\end{equation}
where
\begin{equation}
f'(\theta)=-\frac{ig}{4\pi}\frac{\Gamma^2(1+i\eta) e^{-\pi \eta} }{ 1+igC(k) }\, .
\label{f'}
\end{equation}
The term $f'(\theta)$ is the consequence of the annihilation potential and vanishes if $g\to 0$.
Note that actually $f'$ is angular independent, in accordance with the fact that the zero range potential contributes only in the $s$  partial wave.

As in the case of the positronium spectrum, this scattering amplitude can be compared with the result of the distorted wave Born approximation (DWBA).
The first order DWBA correction to $f_{\cal C}(\theta)$ is given by ~\cite{messiah}
\begin{equation}
\label{dwba}
f^{B} = -\frac{1}{4\pi} \int d\bm{r} \, \psi_{\cal C}(\bm{r},-\bm{k}) igW(\bm{r}) \psi_{\cal C}(\bm{r},\bm{k}),
\end{equation}
where~$\psi_{\cal C}(\bm{r},\bm{k})$ is the scattering Coulomb wave function given by~(\ref{Cwf}).
The action of the pseudo potential $W(\bm{r})$ on the Coulomb wave function can easily be evaluated
\begin{equation}
W(\bm{r}) \psi_{\cal C}(\bm{r},\bm{k}) = \psi_{\cal C}(0,\bm{k}) \delta(\bm{r}) = \Gamma(1+i\eta) e^{-\pi \eta/2} \delta(\bm{r}).
\end{equation}
Thus, the integration in~(\ref{dwba}) results
\begin{eqnarray}
f^B= - \frac{ ig  }{ 4\pi }\Gamma^2(1+ i\eta ) e^{-\pi \eta}.
\end{eqnarray}
This, in fact, coincides with the leading term of  our result of Eq.\ (\ref{f'}) as $g\to 0$.

\section{Annihilation cross section}

The total annihilation cross section accounts for the loss of flux due to annihilation.
This loss is due to the  imaginary absorbing potential.
In the case when $H=\-\Delta_{\bm{r}}+V_1(\bm{r})+ig{V}_2(\bm{r})$, where $V_1(\bm{r})$ and $V_2(\bm{r})$ are  short range potentials,
the scattering wave function
has the asymptotic form
$$
\phi(\bm{r},\bm{k})\sim \exp[i\bm{r}\cdot \bm{k}] + A(\theta)\frac{\exp[ikr]}{r}.
$$
The optical theorem connects the imaginary part of the forward scattering amplitude to the total cross section \cite{messiah}.
If an absorption potential is present, the standard form should be modified as \cite{yak07}
\begin{equation}
\label{opt}
\frac{4\pi}{k}\Im\mbox{m} A(0)  - \sigma   =  \frac{-g}{k} \int d\bm{r} \, {V}_2(\bm{r})|\phi(\bm{r})|^2~,
\end{equation}
where $\sigma$ is the total cross section
\begin{equation}
\label{tot}
\sigma = 2\pi \int d\theta \, \sin\theta |A(\theta)|^2.
\end{equation}
The right-hand side term of~(\ref{opt}) is conventionally interpreted as the absorption cross section
\begin{equation}
\sigma_a=\frac{(-g)}{k} \int d\bm{r} \, {V}_2(\bm{r})|\phi(\bm{r})|^2.
\label{abs-cs}
\end{equation}

In case of long range potentials, when $V_1(\bm{r})$ is the Coulomb potential, the left hand side of the Eq.\ (\ref{opt}) is not well defined. The scattering amplitude in the forward direction is infinite and the integral (\ref{tot}) for the total cross section also diverges. The right hand side should also be redefined if the pseudo potential is used for $V_2$.
In order to obtain a generalization of the optical theorem (\ref{opt}) for Coulomb Hamiltonian with the zero range potential
we consider the partial wave expansion interns of Legendre polynomials $P_\ell(\cos\theta)$
\begin{equation}
f(\theta) = \sum_{\ell=0}^{\infty}(2\ell+1)f_\ell P_\ell(\cos\theta),
\label{f}
\end{equation}
where
\begin{equation}
f_\ell=f^{\cal C}_\ell +f^{'}_\ell .
\label{ff}
\end{equation}
Here the Coulomb partial amplitude reads
\begin{equation}
f^{\cal C}_\ell=\frac{e^{2i\sigma_\ell}-1}{2ik}  \label{fllC}
\end{equation}
with $\sigma_{\ell}=\arg \Gamma(\ell+1+i\eta)$.
The  additional amplitude $f^{'}_\ell$, which is due to the annihilation potential, vanishes for $\ell \ne 0$,
and for $\ell = 0$ reads
\begin{eqnarray}
\label{f'0}
f^{'}_{0}&=&-\frac{ig}{4\pi}\frac{\Gamma^2(1+i\eta) e^{-\pi \eta} }{ 1+igC(k) }\ .
\label{f'ell}
\end{eqnarray}
The problems with the left hand side of the optical theorem (\ref{opt}) in the Coulomb case come from the strong singularity of the Coulomb amplitude (\ref{f_C}) in the forward scattering  direction $\theta=0$. This latter singularity is the consequence of the slow convergence of the series (\ref{f})~\cite{tayl}.
In order to address this problem  let us truncate partial-wave expansions
\begin{equation}
\label{finite}
f_{N}(\theta) = \sum_{\ell=0}^{N}(2\ell+1)f_\ell P_\ell(\cos\theta)
\end{equation}
and
\begin{equation}
\sigma_{N}
= \sum_{\ell=0}^{N} 4\pi (2\ell+1) |f_\ell|^2
\label{tr sigma}
\end{equation}
with~$N$ finite nonnegative integer. Introducing the partial $s$-matrix elements
\begin{equation}
S_{\ell}=1+2ik f_{\ell}
\label{S_ell}
\end{equation}
we write the left-hand side of Eq.\ (\ref{opt}) as
\begin{eqnarray}
\label{rewr}
\frac{ \pi }{ k^2 } \sum_{\ell=0}^{N} (2\ell+1)\left( 1-|S_\ell|^2 \right) 
 =  \frac{4\pi}{k} \Im\mbox{m}\, f_N(0) - \sigma_N.
\end{eqnarray}

If the absorption is absent the scattering is unitary, i.e. $|S_\ell| = 1$, and (\ref{rewr}) yields
\begin{equation}
\label{opt2}
 \frac{4\pi}{k} \Im\mbox{m} \,f_N(0) - \sigma_N  = 0.
\end{equation}
As long as $N$ is finite, this formula is valid for both short range and long range cases. The difference arises  when one is attempting to calculate the limit $N\to \infty$. Indeed, in the short range case, the individual limits of each term in the left hand side of (\ref{opt2}) do exist and that leads to the standard optical theorem.
However, in the long range case, the individual limits do not exist, whereas the collective limit from (\ref{opt2}) is immediate, and the optical theorem takes the form
\begin{equation}
\label{optlim}
\lim_{N\to \infty} \left(\frac{4\pi}{k} \Im\mbox{m}\, f_N(0) - \sigma_N\right)  = 0.
\end{equation}

 Let us consider now the case when besides a Coulomb potential an annihilation potential is also present.
 Since the annihilation effects only the $s$ wave, $|S_{\ell}|=1$ for $\ell \ge 1$, and $|S_{0}|\ne 1$.
More precisely, from Eqs.\ (\ref{ff}-\ref{f'ell}) we have
\begin{eqnarray}
S_0&=&e^{2i\sigma_0}+2ik \, f'_0, \\
S_\ell& =& e^{2i\sigma_\ell}, \ \ \ell \ge 1~.
\end{eqnarray}
This leads  for the left hand side of (\ref{rewr})
\begin{equation}
\frac{ \pi }{ k^2 }\sum_{\ell=0}^{N}  (2\ell+1)\left( 1-|S_\ell|^2 \right) = \frac{ \pi }{ k^2 } \left( 1-|S_0|^2 \right).
\end{equation}
With this equation we obtained a variant of the optical theorem with finite $N$
\begin{equation}
 \frac{4\pi}{k} \Im\mbox{m} f_N(0) - \sigma_N
  = \frac{ \pi }{ k^2 } \left( 1-|S_0|^2 \right).
\end{equation}
However, the right hand side does not depend on $N$, does the $N\to \infty$ limit exists and yields
\begin{equation}
\lim_{N\to\infty} \left(
 \frac{4\pi}{k} \Im\mbox{m} f_N(0) - \sigma_N
 \right)
  = \frac{ \pi }{ k^2 } \left( 1-|S_0|^2 \right).
\label{optCW1}
\end{equation}
As in the case of standard optical theorem (\ref{opt}), the quantity on the right hand side is the measure of the non unitarity
of the $S$ matrix  and in our case it
determines the annihilation cross section
\begin{equation}
\sigma_{\text{ann}} = \frac{ \pi }{ k^2 } \left( 1-|S_0|^2 \right).
\label{sigma_an}
\end{equation}

In order to finalize the generalization (\ref{optCW1}) of the  optical theorem we need to represent the right hand side of (\ref{sigma_an}) in terms of the wave function. This can be accomplished by using the exact representation (\ref{f'0}) for the amplitude $f'_0$ and the properties of the Coulomb amplitudes. After some calculations, which are given in the Appendix,  we arrive at the desired representation
\begin{equation}
1-|S_0|^2 
= \frac{ -g k }{\pi}|\beta|^2,
\label{1-S0}
\end{equation}
which leads to the final form of the annihilation cross section
\begin{equation}
\sigma_{\text{ann}}= \frac{-g}{k}|\beta|^2
\label{sigma_ann}
\end{equation}
and to the final form of the optical theorem
\begin{equation}
\lim_{N\to\infty} \left(
 \frac{4\pi}{k} \Im\mbox{m} f_N(0) - \sigma_N
 \right)
  = \frac{-g}{k}|\beta|^2.
\label{optCW2}
\end{equation}

In previous sections the parameter $\beta$ is represented through the wave function by the integral
\begin{equation}
\beta= \int \mbox{d}\bm{r}\, W(\bm{r})\psi(\bm{r},\bm{k}).
\label{beta int}
\end{equation}
If the value of $\beta$ is calculated from this formula perturbatively, i.e. by using the
 leading term of $\psi$ when $g\to 0$, we get
\begin{equation}
\psi(\bm{r},\bm{k}) = \psi_{\cal C}(\bm{r},\bm{k}) +O(g),
\label{perturb}
\end{equation}
and
\begin{equation}
\beta=\psi_{\cal C}(0,\bm{k}) +O(g).
\label{beta 0}
\end{equation}
Then for the annihilation cross section we have
\begin{equation}
\sigma_{\text{ann}} \sim 
\frac{-g}{k}\int\mbox{d}\bm{r}\,\delta(\bm{r})|\psi_{\cal C}(\bm{r},\bm{k})|^2
\label{Born sigma a}
\end{equation}
and for the optical theorem we get
\begin{equation}
\lim_{N\to\infty} \left(
 \frac{4\pi}{k} \Im\mbox{m} f_N(0) - \sigma_N
 \right)
 \sim \frac{-g}{k}\int\mbox{d}\bm{r}\,\delta(\bm{r})|\psi_{\cal C}(\bm{r},\bm{k})|^2.
\label{optC Born}
\end{equation}

\section{Conclusion}
In this paper we have constructed the zero-range potential for the two particle system with the Coulomb interaction.
The Coulomb potential modifies the asymptotic behavior of the wave function at small distances.
This asymptote has been
studied with the help of the Lippmann-Schwinger integral equation. The pseudo potential
has been constructed such that the wave function obeys the singular asymptotic boundary conditions.
Although, the pseudo potential is constructed for the two body case, its form is universal, and it can be used for several particle systems \cite{mot-yak}.
This work places the delta functional model of the
annihilation potential \cite{iva00} on a sound mathematical basis. The optical theorem, which formally cannot be formulated for charged particles, is reformulated in such a way it becomes applicable to the case of Coulomb plus imaginary absorbing potential of the zero range.
This result facilitates the use of the annihilation
potential for calculating annihilation in collision of positrons with atoms, including rearrangement processes with positronium formation \cite{yak07}.

\begin{acknowledgments}
We would like to thank Prof. Z. Papp for fruitful discussion and  critical reading of the manuscript. This work was partially supported by St Petersburg State University under project No. 11.0.78.2010.
\end{acknowledgments}

\appendix*
\section{Derivation of the representation for the annihilation cross section}
Here we present the derivation of Eq.\  (\ref{1-S0}). Using the exact representations for the Coulomb phase shift and for $f'_0$
we get
\begin{widetext}
\begin{equation}
\label{figbra}
1-|S_0|^2 =  \frac{-g k |\Gamma(1+i\eta)|^2 e^{-\pi\eta} }{\pi |1+igC(k)|^2} 
-\frac{kg^2 |\Gamma(1+i\eta)|^2  e^{-\pi \eta} }{2\pi |1+igC(k)|^2} \left\{ -2\Im\mbox{m}\, C(k) 
 +
\frac{k}{2\pi} |\Gamma(1+i\eta)|^2  e^{-\pi \eta} \right\}
\end{equation}
\end{widetext}
We show that the second term on the right hand side is  zero. Indeed, the first term in figure brackets
can be transformed as
\begin{widetext}
\begin{equation}
-2\Im\mbox{m} \, C(k) 
=  -\frac{k}{2\pi} - \frac{\mbox{n}}{2\pi} \Im\mbox{m}\left[ \log(-2ik) + \psi(1+i\eta) \right] 
  =  -\frac{k}{2\pi} + \frac{\mbox{n}}{4} - \frac{\mbox{n}}{2\pi}\Im\mbox{m}\psi(1+i\eta).
\end{equation}
\end{widetext}
To proceed further, we use the formula~\cite{abram}
\begin{equation}
\Im\mbox{m}\psi(1+iy) = -\frac{1}{2y} + \frac{\pi}{2}\coth(\pi y)
\end{equation}
to get
\begin{equation}
-\frac{k}{2\pi} + \frac{\mbox{n}}{4} - \frac{\mbox{n}}{2\pi}\Im\mbox{m}\psi(1+i\eta)
  =  -\frac{\mbox{n}}{2} \left( \frac{  1 }{e^{2\pi\eta} - 1}\right)~,
\end{equation}
and therefore end up with
\begin{equation}
-2\Im\mbox{m}\, C(k) 
 = -\frac{\mbox{n}}{2} 
\frac{  1 }{e^{2\pi \eta} - 1} 
.
\end{equation}
Now, if we use in the  term
\begin{equation}
\frac{k}{2\pi}|\Gamma(1+i\eta)|^2  e^{-\pi\eta}
\end{equation}
the  relation~\cite{abram}
\begin{equation}
\Gamma(1+iy) \Gamma(1-iy)=\frac{\pi y}{\sinh(\pi y)}
\end{equation}
we get
\begin{equation}
\frac{k}{2\pi}|\Gamma(1+i\eta)|^2  e^{-\pi \eta} = \frac{\mbox{n}}{2} \frac{ 1 }{e^{2\pi\eta} - 1}.
\end{equation}
So, we have shown that the expression in figure brackets in~(\ref{figbra}) equals to zero. Thus,
we arrive at the equality
\begin{equation}
1-|S_0|^2 = \frac{ -gk |\Gamma(1+i\eta)|^2
 e^{-\pi \eta}}{ \pi |1+igC(k)|^2} = \frac{-gk }{\pi}|\beta|^2,
\end{equation}
which proves (\ref{1-S0}).

\bibliography{apstemplate}

\end{document}